\documentclass[superscriptaddress,
amsmath,amssymb,
 aps,
]{revtex4-2}

\usepackage{graphicx}
 \usepackage{hyperref}
\usepackage{xspace}
\usepackage[nolist]{acronym}
\usepackage[usenames,dvipsnames]{xcolor}
\usepackage{booktabs}
 \newcommand{\specialcell}[2][c]{\begin{tabular}[#1]{@{}c@{}}#2\end{tabular}}

\begin{document}
\begin{acronym}
\acro{ADM}{axion dark datter}
\acro{SM}{Standard Model}
\acro{QCD}{quantum chromodynamics}
\acro{PQ}{Peccei-Quinn}
\acro{ALP}{axion-like particle}
\acro{WIMP}{Weakly Interacting Massive Particle}
\acro{PSD}{power spectral densitie}
\acro{DM}{dark matter}
\acro{DFT}{discrete Fourier transform}
\acro{FLL}{flux-lock feedback loop}
\acro{SNR}{signal-to-noise ratio}
\acro{LEE}{look-elsewhere effect}
\acro{TS}{test statistic}
\acro{POM}{polyoxymethylene}
\acro{PTFE}{polytetrafluoroethylene}
\acro{MC}{Monte Carlo}
\acro{AFS}{active feedback stabilization}
\acro{DR}{dilution refrigerator}
\acro{GUT}{grand unified theory}
\acro{SQL}{standard quantum limit}
\acro{BW}{bandwidth}
\acro{Q}{quality factor}
\acro{FOM}{figure of merit}
\acro{SHM}{standard halo model}
\acro{HTS}{high temperature superconducting}
\acro{REBCO}{rare earth barium copper oxides}
\acro{SQUID}{superconducting quantum interference device}
\acro{RMS}{root-mean-square}
\end{acronym}

\newcommand{\SM}{\ac{SM}\xspace}
\newcommand{\PQ}{\ac{PQ}\xspace}
\newcommand{\QCD}{\ac{QCD}\xspace}
\newcommand{\ADM}{\ac{ADM}\xspace}
\newcommand{\DM}{\ac{DM}\xspace}
\newcommand{\ALP}{\ac{ALP}\xspace}
\newcommand{\ALPs}{\ac{ALP}s\xspace}
\newcommand{\WIMP}{\ac{WIMP}\xspace}
\newcommand{\AFS}{\ac{AFS}\xspace}
\newcommand{\PSD}{\ac{PSD}\xspace}
\newcommand{\PSDs}{\acp{PSD}\xspace}
\newcommand{\DFT}{\ac{DFT}\xspace}
\newcommand{\FLL}{\ac{FLL}\xspace}
\newcommand{\SNR}{\ac{SNR}\xspace}
\newcommand{\LEE}{\ac{LEE}\xspace}
\newcommand{\TS}{\ac{TS}\xspace}
\newcommand{\POM}{\ac{POM}\xspace}
\newcommand{\PTFE}{\ac{PTFE}\xspace}
\newcommand{\MC}{\ac{MC}\xspace}
\newcommand{\DR}{\ac{DR}\xspace}
\newcommand{\GUT}{\ac{GUT}\xspace}
\newcommand{\SQL}{\ac{SQL}\xspace}
\newcommand{\Q}{\ac{Q}\xspace}
\newcommand{\FOM}{\ac{FOM}\xspace}
\newcommand{\BW}{bandwidth\xspace}
\newcommand{\MQS}{magnetoquasistatic\xspace}
\newcommand{\SHM}{\ac{SHM}\xspace}
\newcommand{\HTS}{\ac{HTS}\xspace}
\newcommand{\REBCO}{\ac{REBCO}\xspace}
\newcommand{\SQUID}{\ac{SQUID}\xspace}
\newcommand{\SQUIDs}{\acp{SQUID}\xspace}
\newcommand{\RMS}{\ac{RMS}\xspace}

\newcommand{\rhoDM}{\ensuremath{\rho_{\rm DM}}\xspace}
\newcommand{\gagg}{\ensuremath{g_{a\gamma\gamma}}\xspace}
\newcommand{\Jeff}{\ensuremath{\mathbf{J}_{\rm eff}}\xspace}

\newcommand{\bSQL}{beyond-\SQL\xspace}

\newcommand{\DMR}{\mbox{DMRadio}\xspace}
\newcommand{\DMRp}{\mbox{\DMR-Pathfinder}\xspace}
\newcommand{\DMRL}{\mbox{\DMR-50L}\xspace}
\newcommand{\DMRm}{\mbox{\DMR-m$^3$}\xspace}
\newcommand{\DMRGUT}{\mbox{\DMR-GUT}\xspace}
\newcommand{\ABRA}{\mbox{ABRACADABRA}\xspace}
\newcommand{\abra}{\mbox{ABRACADABRA-10\,cm}\xspace}

\newcommand{\fig}[1]{Figure~\ref{fig:#1}}
\newcommand{\sect}[1]{Section~\ref{sec:#1}}
\newcommand{\sects}[2]{Sections~\ref{sec:#1}---\ref{sec:#2}}
\newcommand{\app}[1]{Appendix~\ref{app:#1}}
\newcommand{\eqn}[1]{Equation~(\ref{eqn:#1})}
\newcommand{\tab}[1]{Table~\ref{tab:#1}} 
\title{Introducing DMRadio-GUT, a search for GUT-scale QCD axions}

\author{L.~Brouwer}
\affiliation{Accelerator Technology and Applied Physics Division, Lawrence Berkeley National Laboratory, Berkeley, CA 94720}

\author{S.~Chaudhuri}
\email{sc25@princeton.edu}
\affiliation{Department of Physics, Princeton University, Princeton, NJ 08544}

\author{H.-M.~Cho}
\affiliation{Stanford Linear Accelerator Center, Menlo Park, CA 94025}

\author{J.~Corbin}
\affiliation{Department of Physics, Stanford University, Stanford, CA 94305}

\author{C.~S.~Dawson}
\affiliation{Department of Physics, Stanford University, Stanford, CA 94305}

\author{A.~Droster}
\affiliation{Department of Nuclear Engineering, University of California, Berkeley, Berkeley, CA 94720}

\author{J.~W.~Foster}
\affiliation{Center for Theoretical Physics, Massachusetts Institute of Technology, Cambridge, MA 02139}

\author{J.~T.~Fry}
\affiliation{Laboratory of Nuclear Science, Massachusetts Institute of Technology, Cambridge, MA 02139}

\author{P.~W.~Graham}
\affiliation{Department of Physics, Stanford University, Stanford, CA 94305}

\author{R.~Henning}
\affiliation{Department of Physics and Astronomy, University of North Carolina, Chapel Hill, Chapel Hill, North Carolina, 27599}
\affiliation{Triangle Universities Nuclear Laboratory, Durham, NC 27710}

\author{K.~D.~Irwin}
\affiliation{Department of Physics, Stanford University, Stanford, CA 94305}
\affiliation{Stanford Linear Accelerator Center, Menlo Park, CA 94025}

\author{F.~Kadribasic}
\affiliation{Department of Physics, Stanford University, Stanford, CA 94305}

\author{Y.~Kahn}
\affiliation{Department of Physics, University of Illinois at Urbana-Champaign, Urbana, IL 61801}

\author{A.~Keller}
\affiliation{Department of Nuclear Engineering, University of California, Berkeley, Berkeley, CA 94720}

\author{R.~Kolevatov}
\affiliation{Department of Physics, Princeton University, Princeton, NJ 08544}

\author{S.~Kuenstner}
\affiliation{Department of Physics, Stanford University, Stanford, CA 94305}

\author{A.~F.~Leder}
\affiliation{Department of Nuclear Engineering, University of California, Berkeley, Berkeley, CA 94720}
\affiliation{Physics Division, Lawrence Berkeley National Laboratory, Berkeley, CA 94720}

\author{D.~Li}
\affiliation{Stanford Linear Accelerator Center, Menlo Park, CA 94025}

\author{J.~L.~Ouellet}
\affiliation{Laboratory of Nuclear Science, Massachusetts Institute of Technology, Cambridge, MA 02139}

\author{K.~M.~W.~Pappas}
\affiliation{Laboratory of Nuclear Science, Massachusetts Institute of Technology, Cambridge, MA 02139}

\author{A.~Phipps}
\affiliation{California State University, East Bay, Hayward, CA 94542}

\author{N.~M.~Rapidis}
\affiliation{Department of Physics, Stanford University, Stanford, CA 94305}

\author{B.~R.~Safdi}
\affiliation{Department of Physics, University of California, Berkeley, Berkeley, CA 94720}

\author{C.~P.~Salemi}
\email{salemi@mit.edu}
\affiliation{Laboratory of Nuclear Science, Massachusetts Institute of Technology, Cambridge, MA 02139}

\author{M.~Simanovskaia}
\affiliation{Department of Physics, Stanford University, Stanford, CA 94305}

\author{J.~Singh}
\affiliation{Department of Physics, Stanford University, Stanford, CA 94305}

\author{E.~C.~van~Assendelft}
\affiliation{Department of Physics, Stanford University, Stanford, CA 94305}

\author{K.~van~Bibber}
\affiliation{Department of Nuclear Engineering, University of California, Berkeley, Berkeley, CA 94720}

\author{K.~Wells}
\affiliation{Department of Physics, Stanford University, Stanford, CA 94305}

\author{L.~Winslow}
\affiliation{Laboratory of Nuclear Science, Massachusetts Institute of Technology, Cambridge, MA 02139}

\author{W.~J.~Wisniewski}
\affiliation{Stanford Linear Accelerator Center, Menlo Park, CA 94025}

\author{B.~A.~Young}
\affiliation{Department of Physics, Santa Clara University, Santa Clara, CA 95053} 
\date{\today}

\begin{abstract}
The QCD axion is a leading dark matter candidate that emerges as part of the solution to the strong CP problem in the Standard Model. The coupling of the axion to photons is the most common experimental probe, but much parameter space remains unexplored. The coupling of the QCD axion to the Standard Model scales linearly with the axion mass; therefore, the highly-motivated region $0.4-120\,\textrm{neV}$, corresponding to a GUT-scale axion, is particularly difficult to reach. This paper presents the design requirements for a definitive search for GUT-scale axions and reviews the technological advances needed to enable this program.  
\end{abstract}

\maketitle

\section{Introduction}

There is overwhelming evidence that most of the matter in the Universe is not included in the \SM of particle physics \cite{Bertone:2004pz,Aghanim:2018eyx}.
Despite extensive theoretical and experimental work over the past several decades, the identity of the \DM remains unknown.  One of the leading \DM candidates at present is the quantum chromodynamics (QCD) axion.  The axion is a light boson that was originally postulated as part of the solution to another prominent open question in particle physics, the strong CP problem~\cite{Peccei:1977hh,Peccei:1977ur,Weinberg:1977ma,Wilczek:1977pj}, related to the non-observation of a neutron electric dipole moment.

The QCD axion is the pseudo-Goldstone boson associated with the breaking of a new global $U(1)$ symmetry, called the Peccei-Quinn (PQ) symmetry, in the early Universe~\cite{DiLuzio:2020wdo} at a high energy scale $f_a$.  Due to its coupling to QCD, the axion gains a potential at energies below the QCD confinement scale, leading to the axion having mass $m_a\approx5.7 \,(10^{15}\,{\rm GeV}/f_a)$\,neV~\cite{GrillidiCortona:2015jxo}.  The axion mixes with the neutral pion to acquire a dimension-five coupling with electromagnetism of the form ${\mathcal L} \supset g_{a\gamma\gamma} a {\bf E} \cdot {\bf B}$, with ${\bf E}$ and ${\bf B}$ the electric and magnetic fields, respectively, though ultraviolet contributions to the axion-photon coupling may also be present~\cite{Kim:1979if,Shifman:1979if,Dine:1981rt,Zhitnitsky:1980tq,DiLuzio:2020wdo}.  The axion-photon coupling is $g_{a\gamma\gamma} = C_{a\gamma\gamma} \alpha_{\rm EM} / (2 \pi f_a)$, where $\alpha_{\rm EM}$ is the fine-structure constant and $C_{a \gamma\gamma}$ is a constant typically of order unity that depends on the ultraviolet theory; in the DFSZ model~\cite{Dine:1981rt,Zhitnitsky:1980tq} (KSVZ model~\cite{Kim:1979if,Shifman:1979if}) $C_{a\gamma\gamma} \sim 0.75$ ($C_{a\gamma\gamma} \sim -1.92$), though some axion models achieve  $|C_{a\gamma\gamma}| \gg 1$~\cite{Farina:2016tgd,Agrawal:2017cmd,Sokolov:2021ydn}.  The existence of a nonzero axion-photon coupling is entirely generic due to the low-energy contribution from pion mixing. Therefore, the photon coupling is a promising handle with which to search for the QCD axion.

Theoretical work has shown that axions over a wide range of masses, $m_a\in\left[10^{-12},10^{-2}\right]\,\,$eV, can make up all of the DM~\cite{Dine:1982ah,Preskill:1982cy,abbott1983cosmological,Tegmark:2005dy,Hertzberg:2008wr,Co:2016xti,graham2018stochastic,takahashi2018qcd}. Theoretical priors disfavor masses below this range, which would require super-Planckian $f_a$, as well as higher masses, which are in conflict with stellar cooling observations~\cite{Raffelt:2006cw,Viaux:2013lha,Ayala:2014pea,Carenza:2019pxu,Buschmann:2021juv}.  The preferred mass to allow axions to make up all of the DM depends on the relative timing of PQ symmetry breaking and inflation.  If the PQ symmetry is broken after inflation, the axion mass that yields the observed DM abundance is expected to be $m_a \sim \mathcal{O}(100 \, \mu{\rm eV})$~\cite{Klaer:2017ond,Gorghetto:2020qws,Buschmann:2021sdq}. If instead the PQ symmetry is broken before inflation, then much lighter axion masses are possible through mechanisms including anthropic selection~\cite{Tegmark:2005dy,Hertzberg:2008wr}, late-time entropy dilution~\cite{Co:2016xti}, and inflationary dynamics~\cite{graham2018stochastic,takahashi2018qcd}. A particularly well-motivated part of this lower mass range is $m_a\sim1-100$\,neV, which corresponds to $f_a$ near the \GUT scale.  Axion decay constants $f_a \sim 10^{14} - 10^{16}$\,GeV arise naturally in the context of String Theory constructions (see, {\it e.g.},~\cite{Green:1984sg,Svrcek:2006yi,Conlon:2006tq,Acharya:2010zx,Ringwald:2012cu,Cicoli:2012sz,Halverson:2019cmy}), which generically produce PQ symmetries through the compactification of the extra dimensions~\cite{Witten:1984dg}, and \GUT field theories~\cite{Wise:1981ry,Ballesteros:2016xej,Ernst:2018bib,DiLuzio:2018gqe,Ernst:2018rod,FileviezPerez:2019fku,FileviezPerez:2019ssf,Co:2016xti}.  At these low masses, the axion Compton wavelength in our galaxy is large relative to the size of a detector, which has motivated experiments that use lumped-element circuits to detect axion-photon interactions~\cite{Sikivie:2013laa,Chaudhuri:2014dla,Kahn:2016aff,silva2016design}.  This detection technique exploits axion interactions with a static magnetic field by inductively coupling the resulting effective current into an LC circuit.  Three first-generation lumped-element experiments, \abra, SHAFT, and ADMX-SLIC, have recently set world-leading limits \cite{Ouellet:2018beu,Salemi:2021gck, Gramolin:2020ict, Crisosto:2019fcj}.

Current results with this method are several orders of magnitude less sensitive than would be necessary to detect QCD axions at neV masses, but there are efforts underway to build more sensitive lumped-element detectors. \DMR is a program of experiments that aims to incrementally improve the sensitivity of the lumped-element technique \cite{ouelletprobing}.  The first of these experiments is \DMRL, which aims to search for axion-like particles over a wide range of masses with couplings $\gagg<10^{-14}\,\textrm{GeV}^{-1}$ and is currently being constructed.  A larger, more sensitive detector \DMRm will probe $\sim120-800\,\textrm{neV}$ axions at DFSZ sensitivities \cite{dine1981simple,zhitnitskij1980possible}.

This paper presents the path toward detecting QCD axions at GUT-motivated masses. Achieving the required sensitivity necessitates improvements in several of the core components of the experiment: magnet, cryogenics, amplifier, and resonator. We begin in \sect{lumpedelement} by describing the lumped-element detection method and presenting the scan rate, which we use to set the performance required for these core components. \sect{magcryo} discusses the requirements on the magnet and cryogenic systems and potential paths for improvement in these areas. \sect{amps} presents the need for beyond-\SQL amplifiers, and \sect{resonator} describes improvements that can be made to the tunable resonator. We conclude in \sect{reach} by comparing different experiment configurations that could definitively search for QCD axions in the $0.4-120\,\textrm{neV}$ mass range (corresponding to rest-mass frequencies $\nu_{a}=m_{a}c^{2}/h$ of $100\,\textrm{kHz}-30\,\textrm{MHz}$), probing masses associated with the GUT scale.

\section{lumped-element detection}\label{sec:lumpedelement}

The axion is a type of classical wave-like DM because its small mass and consequently large number density lead to high occupation numbers per quantum state.  This is in contrast to the particle-like interactions probed in searches for heavier DM candidates \cite{Akerib2020,XENON1T,Agnese2018}.  At $m_a\sim1\,\rm{neV}$, the axion's Compton wavelength is $\sim1\,\rm{km}$, which is much larger than any current or planned DM experiment. In this regime, the \MQS approximation holds, and the variation in the electric field associated with the axion-photon interaction is small. The axion-photon interaction can then be interpreted in terms of an axion-modified Amp\`ere's law:
\begin{equation}
    \nabla\times\mathbf{B}=-\gagg \sqrt{\frac{\hbar \epsilon_{0}}{c}} \left(\mathbf{E}\times\nabla a-\frac{\partial a}{\partial t}\mathbf{B}\right)
\end{equation}
where $a$ is the axion field. Here, we have neglected to include the displacement current term, $\partial\mathbf{E}/\partial t=0$, (being in the \MQS limit) or other current sources \cite{Kahn:2016aff,Ouellet:2018nfr}.  In the \SHM \cite{turner1990periodic,freese2013colloquium}, the spatial gradient of the axion is expected to be small relative to its time derivative, and so we take as our signal only the second term on the right hand side, which can be written as an effective current that is proportional to the magnetic field that sources the axion interactions:
\begin{equation}\label{eq:Jeff}
    \textbf{J}_{eff}=\frac{\sqrt{\hbar c}}{\mu_0}\gagg\sqrt{2\rho_{DM}}\cos(m_at)\textbf{B} \,,
\end{equation}
where $\rho_{DM}\approx0.45\,\textrm{GeV/cm}^3$ is the local energy density of axion DM~\cite{de2021dark}.

\begin{figure}[h]
    \centering
    \includegraphics[width=0.7\textwidth]{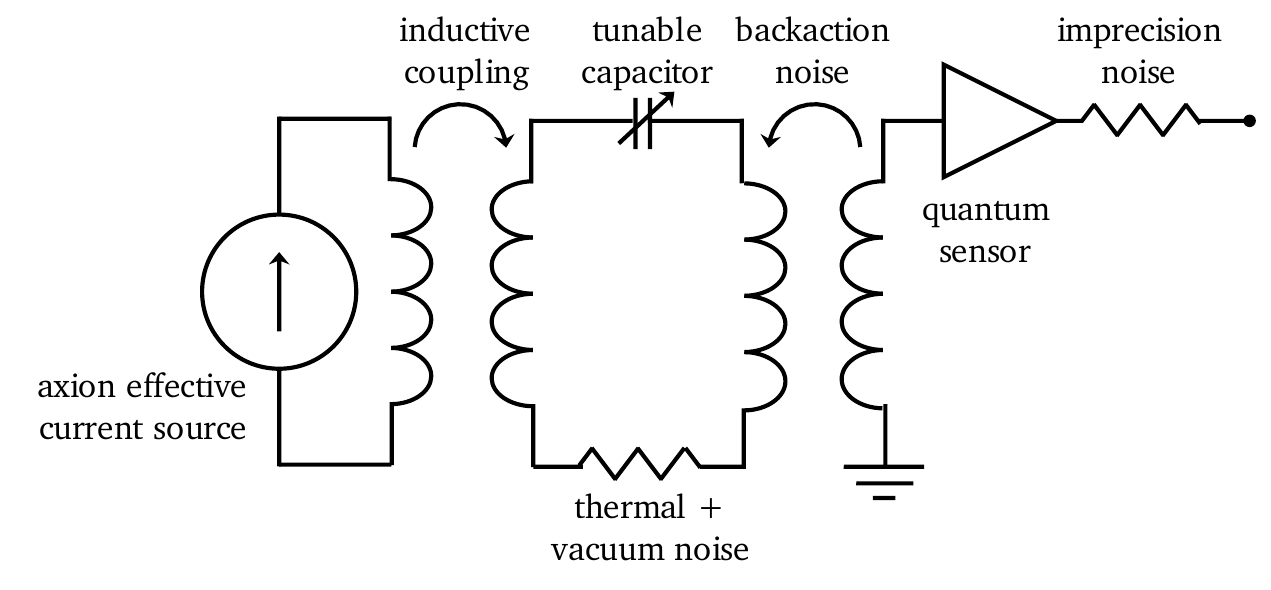}
    \caption{Effective circuit diagram for a lumped-element axion experiment. On the left, the axions' interaction with a strong magnetic field creates an effective current source that can be inductively coupled to a superconducting circuit.  The central loop acts as an LC resonator with quality factor $Q$, which can enhance signals that match the resonance frequency.  To scan over a range of axion masses, the capacitor is tuned.  Finally, the signal is read out using high-precision quantum sensors and analyzed offline.  Noise enters the system in the resonator as thermal and vacuum noise and on the readout as amplifier-added noise.}
    \label{fig:circuit}
\end{figure}

Using the \MQS approximation, the axion effective current, $\textbf{J}_{eff}$, produces oscillating magnetic fields that can be inductively coupled to a detection circuit consisting of inductors and capacitors~\cite{Kahn:2016aff}, as shown in \fig{circuit}. This is the lumped-element circuit model from which the detection method takes its name. The signal from an axion of mass $m_a$ is resonantly enhanced if its corresponding frequency is close to the circuit's LC resonance.  Because we do not know $m_a$ \emph{a priori}, the LC resonance must be tuned over a wide range of frequencies.  The time spent at each frequency determines the experiment's scan rate.

\subsection{Sensitivity}\label{sec:sensitivity}

The overall sensitivity of axion experiments can be evaluated in terms of their scan rate, which quantifies how quickly a detector can be sensitive to a given value of \gagg over a range of axion masses. The rate at which the resonance frequency, $\nu_r$, is tuned is (see Appendix \ref{app:scanRateDeriv} for a derivation)

\begin{align}\label{eqn:scanRate}
    \frac{d\nu_r}{dt} \approx 41 \frac{\textrm{kHz}}{\textrm{year}} & \left( \frac{3}{\textrm{SNR}} \right)^{2} \left( \frac{g_{a\gamma\gamma}}{10^{-19}\ \textrm{GeV}^{-1}} \right)^{4} \left( \frac{\rho_{\rm DM}}{0.45\ \textrm{GeV}/\textrm{cm}^{3}} \right)^{2} \nonumber\\
    &\times\left( \frac{\nu_{r}}{100\ \textrm{kHz}} \right)\left( \frac{c_{PU}}{0.1} \right)^{4}  \left( \frac{B_{0}}{16\ \textrm{T}} \right)^{4} \left( \frac{V}{10\ \textrm{m}^{3}} \right)^{10/3} \left( \frac{Q}{2 \times 10^{7}} \right) \left( \frac{10\ \textrm{mK}}{T} \right) \left( \frac{0.1}{\eta_{A}} \right)  \,.
\end{align}
This scan rate depends on the desired \SNR on resonance for a given coupling, \gagg, as well as the bandwidth over which that \SNR is approximately constant, which we will refer to here as the sensitivity bandwidth. The \SNR essentially determines how long the experiment must integrate at each frequency step, and the sensitivity bandwidth sets the minimum spacing between steps.

The first line in \eqn{scanRate} contains the axion and DM physics, where we have assumed that the DM distribution follows the \SHM \cite{freese2013colloquium}.  The second line describes experimental parameters: $c_{PU}^2$ is a proxy for the ratio of energy that is coupled into the resonant circuit to energy stored in axion-generated magnetic field.  It is analogous to the form factor in a cavity axion experiment \cite{sikivie1985detection}. A typical value is $c_{PU} \sim 0.1$ \cite{Chaudhuri:2021xjd}.  $B_0$ is the peak strength of the DC magnetic field assuming a toroidal magnet with a ratio of outer to inner radii $r_{out}/r_{in}=2$.  $V$ is the total volume of the pickup structure; see the appendix for a discussion on the parametric dependence of scan rate on volume. $Q$ is the quality factor of the resonant circuit. The last two terms represent thermal noise (scaling with temperature $T$) and amplifier noise ($\eta_{A}$).  Note that we have assumed that thermal noise dominates over amplifier noise, as is true at the low frequencies this method searches.

As evident from \eqn{scanRate}, the most effective ways of increasing the scan rate and hence overall sensitivity are to increase the magnetic field strength and volume as well as the detector-axion coupling, $c_{PU}$, which scale with high powers.  However, as we discuss below, order(s) of magnitude improvement may also be made to $Q$ and $\eta_A$.  The developing technologies that can facilitate these improvements are discussed in \sects{magcryo}{resonator} and summarized in \tab{techTable}.

\begin{table}[h]
    \centering
    \begin{tabular}{ccc} \toprule
        {\textbf{Technology}} & {\textbf{Parameter(s) affected}} & {\textbf{Target value}} \\ \midrule
        \specialcell{\textbf{Magnets}\\\specialcell{REBCO magnets\\Nb$_3$Sn magnets}}  & $B_0,V$ & 16\,T, 10\,m$^3$ \\[0.5cm]
        \specialcell{\textbf{Backaction evasion}\\RF Quantum Upconverters} & $\eta_A$ & -20\,dB \\[0.3cm]
        \specialcell{\textbf{Resonators}\\\specialcell{Passive resonators\\Active feedback}}  & $Q$ & $20\times10^6$ \\
        \bottomrule
    \end{tabular}
    \caption{R\&D thrusts for improving the sensitivity of future lumped element experiments.  The target values shown are those we assume for our baseline scenario of a 6.2\,year scan.}
    \label{tab:techTable}
\end{table}

\subsection{Predecessor experiments}

The lumped-element technique for axion detection has been demonstrated by several experiments.  The first experiment to set limits using this technique was \abra, using a small 1\,T toroidal magnet coupled to a broadband readout circuit \cite{Ouellet2019a,Ouellet2019b,Salemi2021}. The SHAFT experiment used ferromagnets in a double-toroid configuration to also set limits in the mass range near 1\,neV \cite{Gramolin2021}.

The experiment ADMX-SLIC used a different configuration to look for axions at slightly higher masses \cite{Crisosto2020}.  A solenoidal magnet provided a strong field (up to 7\,T) for the axions to couple to, and the resulting signal was read out via a tunable resonant circuit.  The \DMRp dark photon search also demonstrated the use of a tunable resonant readout, although without the magnetic field that is required for axion searches  \cite{Phipps2020}.

Using this lumped-element method, each of these small experiments was able to search for axions in unexplored \gagg~parameter space.  They demonstrated long-term operation and data-taking stability, successful resonator tuning at cryogenic temperatures, and the ability to integrate a tesla-scale magnetic field in conjunction with a sensitive superconducting readout. Upcoming experiments will address the challenges encountered by these prototype detectors, including the isolation of vibrational and other environmental noise sources and the development and maintenance of a high-Q tunable resonator.

The central objective for future experiments is improving sensitivity to \gagg as dictated by the scan rate in \eqn{scanRate}.  The next experiment to use the lumped-element technique, \DMRL, focuses on improving the axion-detector coupling, increasing the science volume, and implementing a resonant readout.  \DMRL will have a toroidal $\sim50\,$L magnetic field volume, up from the $\sim1$\,L volume of \abra, and will significantly improve the coupling factor by enclosing the axion effective current in a superconducting pickup sheath.  This new pickup will also allow the experiment to use a resonator with a Q-factor of $\sim10^{6}$ \cite{falferi1994high,nagahama2016highly}.

The next experiment in the DMRadio program will scale the technique to 1\,$\textrm{m}^3$. At this scale, a definitive search at slightly higher masses is possible. \DMRm is designed to search for masses $m_a\in120-800\,\textrm{neV}$, corresponding to frequencies $\sim30-200\,\textrm{MHz}$ \cite{m3inprep}. This higher frequency range has interesting challenges inherent to approaching the breakdown of the \MQS approximation. 

The \DMRL and \DMRm experiments will be able to provide insight into the potential challenges in construction, integration, and operation of \DMRGUT at a larger scale than was achieved by the prototype experiments.  For example, we expect to learn more about the feasibility of cooling and operating a massive magnet in tandem with highly sensitive superconducting electronics.  \DMRGUT will combine these lessons with upcoming technological advances in order to achieve the exquisite sensitivities necessary for a successful GUT-scale experiment.  In the following sections we lay out the R\&D path towards this goal.

\section{Magnet and cryogenics}\label{sec:magcryo}

The predecessor experiments use one of two basic geometries: a toroidal or solenoidal magnet. The toroidal design has some major advantages. In a toroid, the magnetic field is contained within the coils, unlike a solenoidal field with field lines that return outside of the magnet bore.  Magnet quenches are thus much less dangerous and also pose minimal threat to the superconducting electronics, which need to operate in a low-field region.  Contained fields also allow the pickup structure to naturally sit in a field-free location.  This placement avoids introducing leading-order magnet-related backgrounds into the readout, and it also allows the pickup to be easily constructed from superconducting materials, leading to higher resonator $Q$'s.  However, the pickup geometries used in toroidal designs suffer from parasitic resonances (due to capacitance in overlapping sheath components), which can limit their sensitivity at high frequencies. Because \DMRGUT is designed for low frequency, the issue of parasitic resonances for a toroidal architecture is mitigated. 

In the solenoidal design, the pickup is most sensitive when inside the high field region. However, this placement increases the risk of magnet-sourced noise, and large fields make the construction of a high-$Q$ circuit difficult \cite{ulmer2009quality}.  High-field solenoids are produced for many applications, so they benefit from existing designs and infrastructure for their construction. They are thus generally easier and cheaper to build.

Experience from the toroidal \DMRL and solenoidal \DMRm experiments will be important inputs into the decision of \DMRGUT's geometry.  For example, magnetic pinning effects in the solenoidal field could reduce vibrational noise, making that design more competitive, especially if vibrations are a dominant noise source \cite{Brubaker:2017ohw}. Simulations and measurements of the \DMRm detector will guide our estimates of the importance of vibrational backgrounds in a potential solenoidal \DMRGUT.  Likewise, we will be able to measure any stray fringe fields from the \DMRL toroid that can produce interference and vibrational noise in the pickup and use that to gauge tolerances on the construction of a \DMRGUT toroid.  Ultimately, we expect that in a solenoid, the disadvantages of the pickup being in the high-field region would be a limitation, making the toroidal design the most likely outcome.

Regardless of the geometry, the \DMRGUT magnet will have to be large and thus heavy. As in \DMRL and \DMRm, we plan to cool only the resonator to sub-K temperatures. However, even with the goal of cooling the magnet only to its superconducting transition temperature, $T_c$, careful consideration will have to be made in the design of the cryogenics. For the toroidal design, some of the cryogenic constraints are mitigated by surrounding the magnet with a superconducting sheath, as in \DMRL, which ensures that the resonator does not couple to magnet loss or thermal noise that would otherwise degrade sensitivity. The primary decision that will determine the cryogenic requirements is the choice of magnet wire materials, which can be grouped into two categories, traditional, low-$T_c$ superconductors and newer, high-$T_c$ superconductors such as \REBCO.

\subsection{Low-$T_c$ magnet}

There are many low-$T_c$ superconductors, but generally Nb-based compounds are favored for the construction of high field magnets.  Here we explore two common choices that could be used in \DMRGUT.

The first option is to use a traditional NbTi superconducting magnet, as has been done for the other lumped-element detectors.  A NbTi-based design would require cooling the magnet below 10\,K and would also limit the peak magnetic field strength to $\sim$10\,T \cite{godeke2007limits}, posing stringent practical constraints on \DMRGUT.  In order to maintain our goal sensitivity, this restriction on field strength would require greatly increasing the magnet volume, which would further increase the load on the cryogenics and/or require large improvements in the noise and quality factor of the receiver.

A second option is to use a Nb$_3$Sn magnet. While more costly and difficult to use in magnet fabrication, Nb$_3$Sn can produce the benchmark 16\,T magnetic field for a GUT-scale QCD axion search. Significant work has been done in related fields to optimize use of Nb$_3$Sn magnets in large-scale physics experiments.  For example, the ITER fusion project will utilize solenoidal and toroidal Nb$_3$Sn coils operating at a temperature of $\sim$5 K with peak fields of 13 T and 11.8 T, respectively \cite{mitchell2008iter}. The latter toroidal magnets, used for confining the plasma, will have 8500\,m$^3$ volume and 51\,GJ of stored energy.  This material will also be used as part of the High-Luminosity LHC, for the accelerator's dipole and quadrupole magnets, reaching 12\,T peak fields and 2.5 GJ of stored energy \cite{bottura2012advanced,todesco2018progress}.

\subsection{High-$T_c$ magnet}

Additional options beyond low-$T_c$ magnets are on the horizon due to recent and emerging advancements in \HTS materials. \HTS-based magnets have several advantages over traditional niobium-alloy-based magnets. First, because of the higher $T_c$, the cooling requirements for the large magnet mass can be relaxed; \HTS magnets can operate at a few tens of kelvin instead of a few kelvin. Second, \HTS materials such as \REBCO are able to remain superconducting in fields exceeding 35\,T \cite{maeda2013recent}, allowing us to increase $B_0$ and provide additional engineering margin for a GUT-scale search.

\HTS magnets are already being used for a variety of projects, most notably in fusion, which, similar to \DMRGUT, requires high fields and large volumes.  One such project is the SPARC tokomak \cite{Creely2020}, a toroidal magnet made of YBCO tape with an average magnetic field of 12.2\,T and a major radius of 1.85\,m for a total stored energy of 110\,MJ.  Recently, the SPARC effort demonstrated peak fields of 20\,T.

By taking advantage of these developments in magnet technology, lumped-element experiments can greatly improve their reach.  \DMRGUT plans to achieve a 12\,T \RMS field (equivalent to a 16\,T peak field for a toroid with the same approximate aspect ratio as \DMRL) in a 10\,m$^3$ volume, for a total stored energy of 573\,MJ, well within the bounds of current and planned magnets.  However, even with these magnet parameters, running the experiment with \SQL amplifiers would require a runtime of $\sim50\,$years.  A powerful magnet will consequently need to be paired with beyond-\SQL amplifiers, as we explore in the next section.

\section{Beyond-SQL amplifiers}\label{sec:amps}

Simultaneous with the explosion in interest in wave-like axion DM, there has been a revolution in quantum sensing, enabling measurements at unprecedented sensitivity. These new capabilities will facilitate searches for QCD axions over a much wider parameter space.

There are two primary requirements on quantum sensors for axion searches.  First, they must be sensitive to tiny signals, since a receiver couples a miniscule amount of power from the axion field \cite{Chaudhuri:2021xjd} (e.g., $\lesssim 10^{-22}\,\textrm{W}$ for traditional cavity-based DFSZ axion searches in several-tesla magnetic fields \cite{Brubaker:2017ohw}).  Second, because the axion mass is unknown, measurements must be able to scan over orders of magnitude in frequency.  Thus, usable sensors must not only enable a high \SNR, but they must also be able to do so over a wide bandwidth in frequency in order to increase step sizes and achieve a fast scan.  As described in ref. \cite{Chaudhuri:2021xjd} (see also Section VI B of \cite{tome} and \cite{Chaudhuri:2019ntz} for an extended discussion), the figure of merit for readout sensors is thus a frequency-integrated sensitivity, increasing with both the on-resonance \SNR and the bandwidth over which that \SNR can be maintained.

Progress has already been made towards implementing beyond-\SQL measurement in axion searches in the microwave frequency range ($\gtrsim 1$ GHz). The HAYSTAC experiment has demonstrated an axion receiver in which the cavity is injected with squeezed electromagnetic vacuum, doubling the \BW and consequently the experiment's scan rate over previous work \cite{backes2021quantum}. Recent work on qubit-based photon counting has demonstrated a noise reduction that could yield a factor of 1,300 increase in scan rate compared to a quantum-limited search \cite{dixit2021searching}.

\begin{figure}[h]
    \centering
\includegraphics[width=0.7\textwidth]{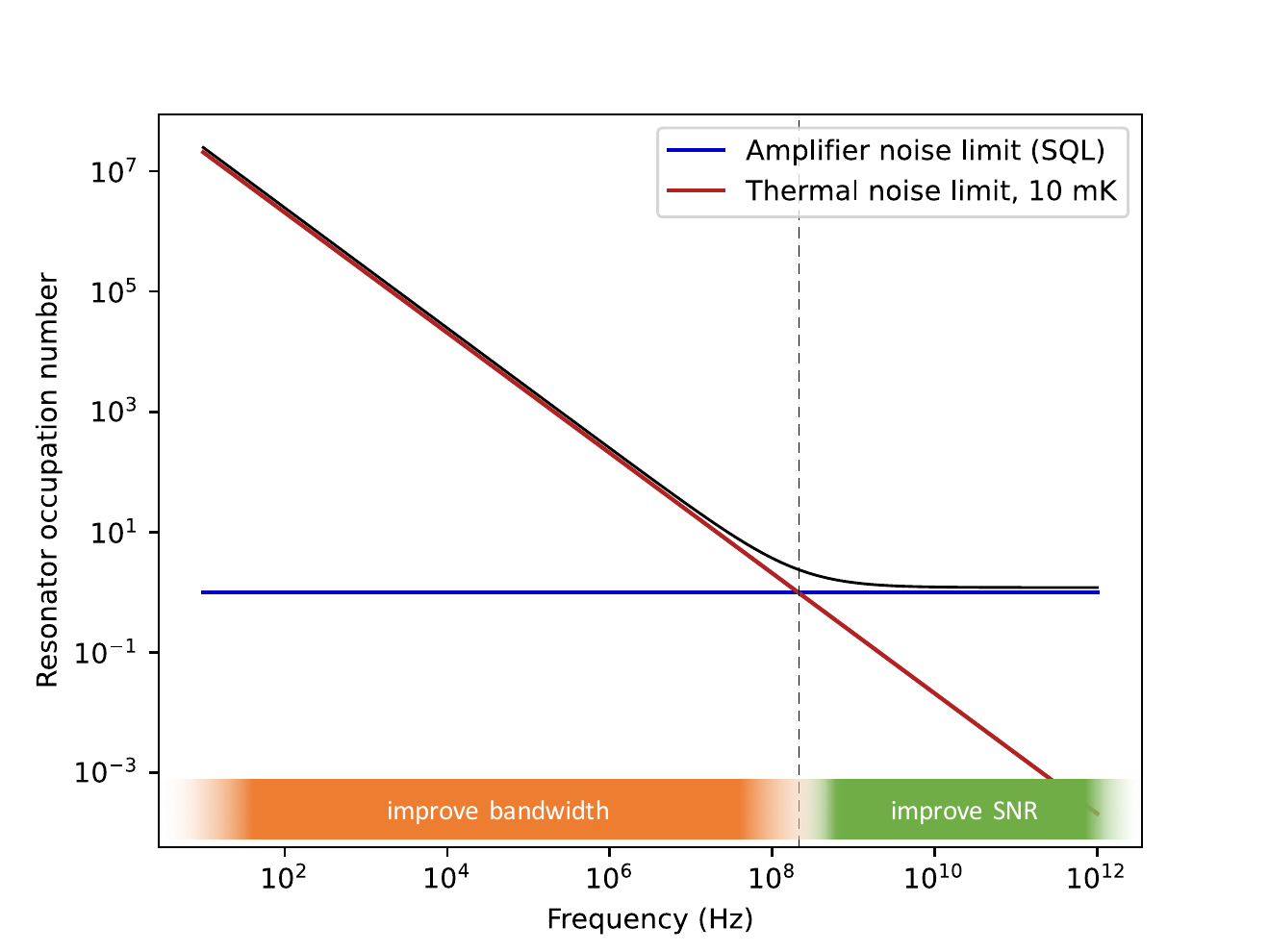}
    \caption{There are two distinct noise regimes in axion searches.  At high frequencies, thermal noise is subdominant to readout noise, and so improvements in readout performance beyond the \SQL directly improve the \SNR.  At low frequencies, thermal noise is dominant, and so beyond-\SQL amplification does not directly improve the \SNR.  Instead, integrated sensitivity can be improved by increasing the sensitivity \BW.}
    \label{fig:detectionRegimes}
\end{figure}

To enable DMRadio-GUT, we must develop quantum sensing technologies at much lower frequencies, in the range of 1\,kHz-100\,MHz. Unlike the upper end of the microwave frequency range (the green region in \fig{detectionRegimes}), at the frequencies of a GUT-scale axion, the resonator is thermally occupied, which means that quantum noise-evading techniques such as squeezing or photon counting do not significantly improve \SNR relative to a quantum-limited readout. The objective in this range (\fig{detectionRegimes}'s orange region) is thus to improve the frequency-integrated sensitivity by executing quantum protocols on thermal states to increase the bandwidth rather than the \SNR \cite{caves1982quantum,clerk2010introduction}.

At our low frequencies, with a quantum-limited amplifier, resonator thermal noise dominates quantum noise over a broad range of frequencies much larger than the resonator bandwidth. Over this range of thermal noise domination, the \SNR is approximately constant because the resonator response to the signal and noise are identical---they both vary with frequency approximately as Lorentzian functions.  This range over which the \SNR is constant is what we call the `sensitivity~bandwidth,' because we are equally sensitive to an axion signal throughout, even in the region outside the resonator bandwidth.  To enhance scan sensitivity with a quantum sensing protocol, it is then natural to increase the sensitivity bandwidth.

\begin{figure}[h]
    \centering
    \includegraphics[width=0.49\textwidth]{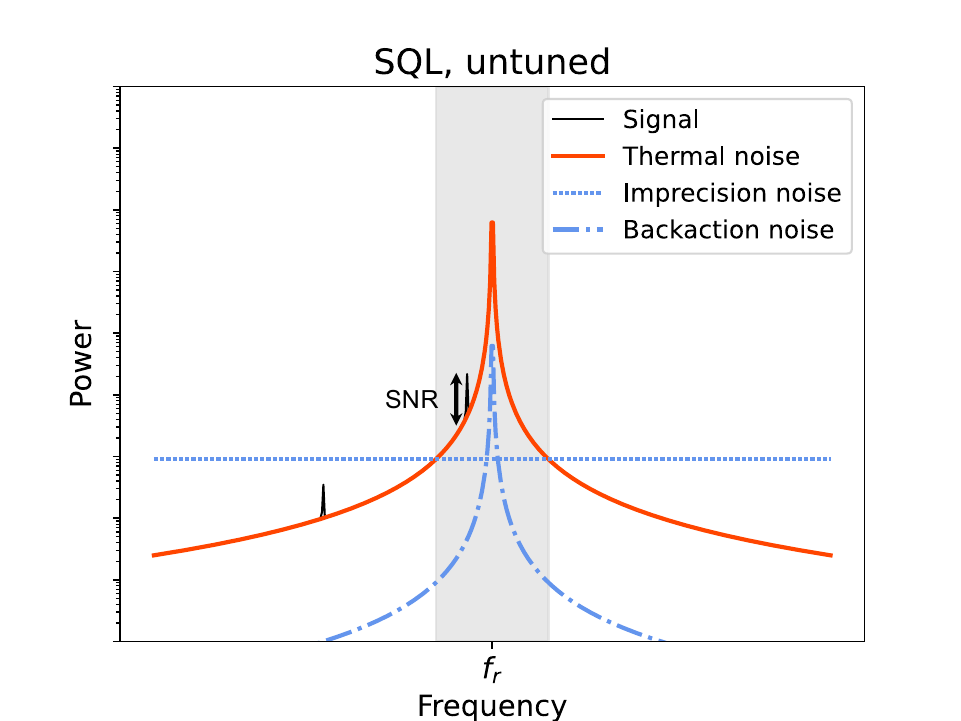}
    \includegraphics[width=0.49\textwidth]{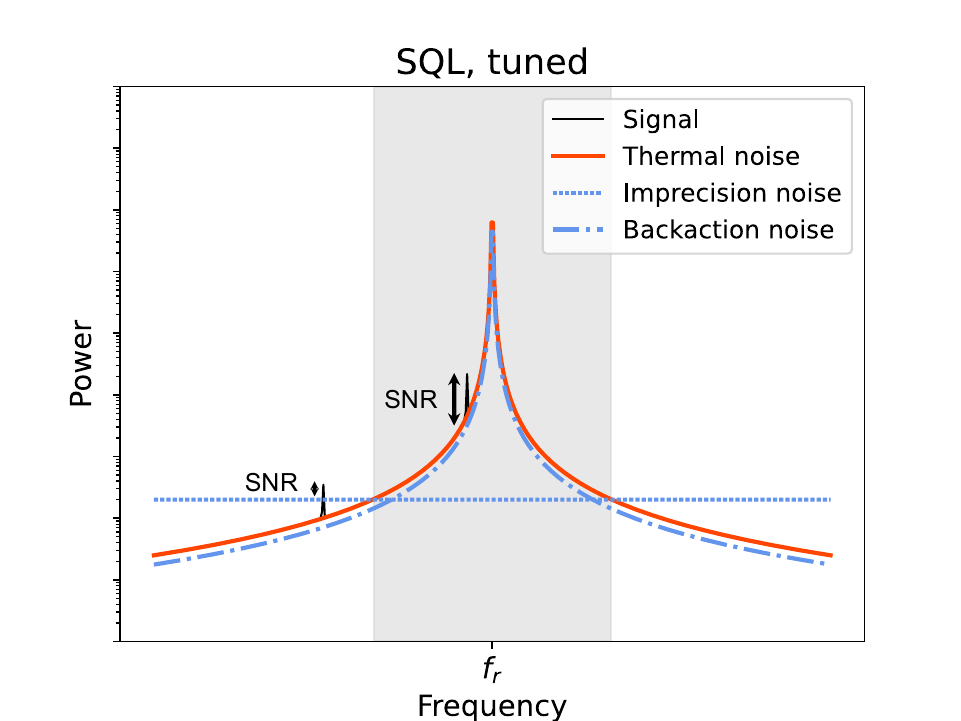}\\
    \includegraphics[width=0.49\textwidth]{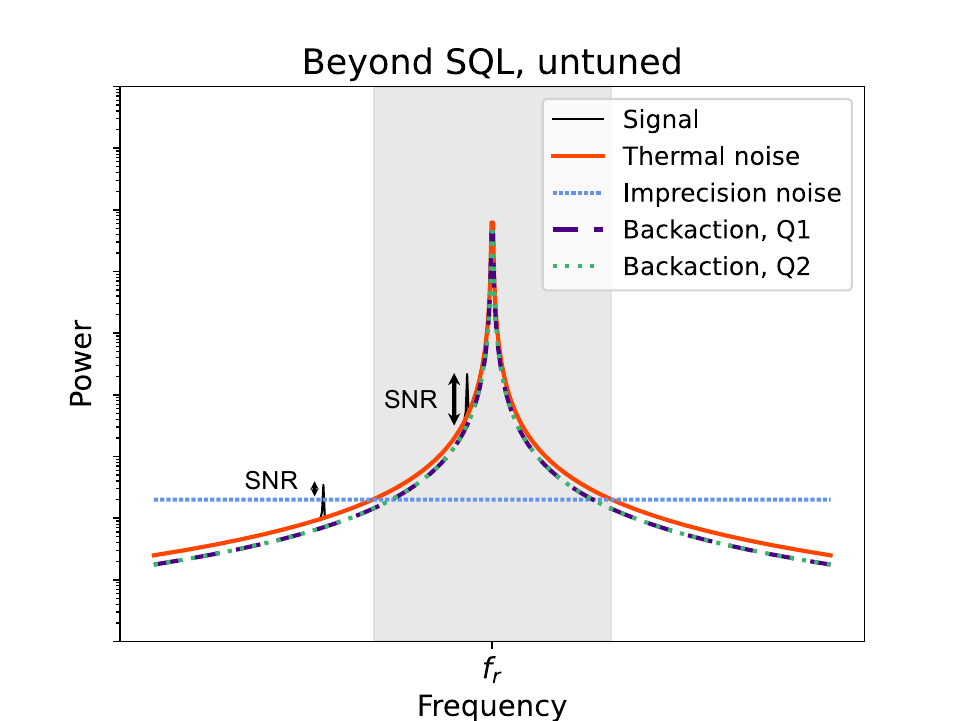}
    \includegraphics[width=0.49\textwidth]{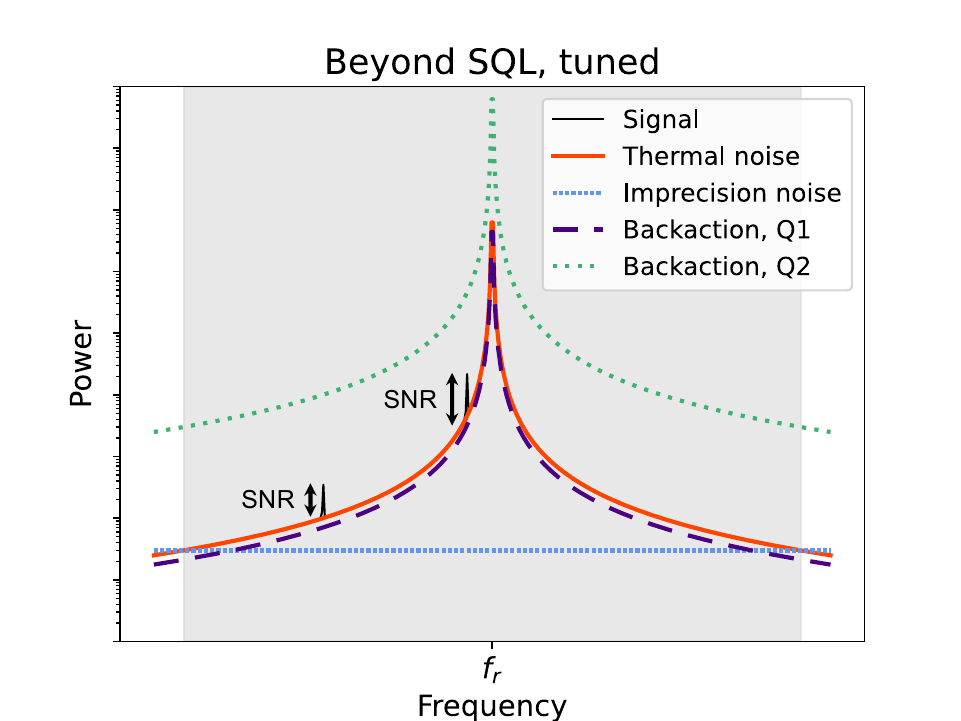}
    \caption{Optimizing the sensitivity bandwidth for quantum-limited amplification, \emph{top}, and beyond-\SQL amplification via backaction evasion, \emph{bottom}.  The cases with untuned amplifier input coupling are on the left, and the more optimally coupled cases are shown on the right. Note that the \SNR (labeled in black) is similar for the near-resonance signal in each case while the sensitivity \BW (grey band) can be greatly increased by tuning the coupling.  A second, off-resonance signal is shown, buried under the noise for an untuned, \SQL setup.  By tuning the coupling, the signal may rise above the amplifier noise, and in the final, tuned \bSQL case, it can reach its maximum \SNR above the thermal noise background.  Using \bSQL amplifiers allows the experimenter to shift more of the imprecision noise into backaction without affecting the \SNR of signals in the sensitivity bandwidth; the noise only goes into the unmeasured quadrature, here Q2.}
    \label{fig:biggerBW2}
\end{figure}

As illustrated in Fig. \ref{fig:biggerBW2}, a backaction evasion scheme both reduces the minimum added noise of the readout below the one-photon SQL and increases the sensitivity bandwidth. The \BW improvement relies on the fact that the amplifier noise can be divided into two components with different inherent line shapes; see Appendix \ref{app:scanRateDeriv}, Section V of \cite{clerk2010introduction}, and Appendix F of \cite{tome} for more information. The first component, known as imprecision noise, is a flat background that is added on the amplifier output.  The second component is backaction, which is injected back into the detector resonator through the amplifier input.  Because it goes through the resonator, backaction shares the resonator's Lorentzian line shape.  For an example of how this works, we can examine DC~\SQUIDs, which have demonstrated noise as low as 20 times the quantum limit in commercial implementations \cite{falferi200810}, and which are the baseline for \DMRL and \DMRm.  In \SQUIDs, the imprecision noise consists of intrinsic voltage fluctuations on the amplifier output, and the backaction noise consists of circulating currents in the SQUID loop, which induce a voltage and thus resonantly enhanced currents in the input coil \cite{clarke2006squid,clarke1979optimization}. 

The sum of the the imprecision and backaction components is the total amplifier noise plotted in \fig{detectionRegimes}, and it is this sum that is subject to the \SQL.  By tuning the strength of the input coupling of the amplifier, one can adjust the balance of imprecision versus backaction noise.  Improvement on this scheme can be accomplished with backaction evasion, shown in the bottom row of \fig{detectionRegimes}, where the backaction in one quadrature of the signal is reduced at the expense of increased backaction in the other quadrature. As in the top row, by coupling the amplifier more strongly to the readout, the imprecision noise is reduced, increasing backaction noise---but now only in the unmeasured quadrature.  This process thus maintains the noise levels of the measured backaction quadrature and lowers the imprecision noise, increasing the sensitivity bandwidth.

An alternative approach to backaction evasion is variational readout. For amplifiers subject to the \SQL, the imprecision and backaction noise are uncorrelated. By introducing correlations between the imprecision and backaction, one can achieve a better noise match and reduce the total noise off-resonance, thereby increasing the sensitivity bandwidth. The introduction of correlations to reduce off-resonance noise has been demonstrated in microwave measurements of mechanical resonators \cite{kampel2017improving}.

One possibility for implementing these quantum protocols for 1\,kHz-100\,MHz electromagnetic signals is the RF Quantum Upconverter (RQU); see also Chapters 6 and 7 of \cite{chaudhuri2019dark} as well as \cite{RQUinprep1} and \cite{RQUinprep2}. Like a DC~\SQUID, the RQU takes a flux signal as its input, which facilitates its integration as an upgrade to the DMRadio readout chain. However, in contrast to a DC~\SQUID, for which the input and output signals match in frequency, the RQU upconverts kHz-MHz flux signals to GHz voltage signals, allowing us to leverage existing microwave quantum technologies \cite{krantz2019quantum}. The RQU is realized by embedding an interferometer of Josephson junctions in a superconducting microwave resonator.  This setup implements a measurement Hamiltonian analogous to that of cavity optomechanics, where decades of work has enabled quantum sensing on low-frequency, mechanical systems (e.g., LIGO) \cite{abbott2009observation,aspelmeyer2014cavity}.  RQUs can thus use analogous protocols for quantum noise-evasion, such as the backaction evasion scheme proposed for mechanical resonators in \cite{clerk2008back}; in an RQU, this scheme may be executed by exciting the microwave resonator with an amplitude-modulated drive.

By implementing quantum protocols, we aim to achieve readout noise levels of $\eta_{A}=0.1$, corresponding to 20\,dB of backaction-noise reduction and a factor of ten increase in scan rate. Quantum squeezing of 10\,dB has already been achieved in the GHz regime using Josephson parametric amplifiers, and more recently, quantum squeezing of 20\,dB (analogous to our goal) was demonstrated in a spin ensemble controlled by optical light \cite{castellanos2008amplification,hosten2016measurement}. As part of demonstrations toward \DMRGUT, RQUs and quantum protocols may be integrated into the \DMRL testbed.

\section{Resonator R\&D}\label{sec:resonator}

To probe GUT-scale QCD axion DM, it will be necessary to improve lumped-element resonators in the frequency range 1\,kHz-100\,MHz. We envision a two-pronged approach: 1) increasing the quality factor of a passive resonator and 2) implementing active matching to increase bandwidth without reducing resonator gain.

\subsection{Passive resonator}

State of the art lumped-element resonators have demonstrated quality factors of $\mathcal{O}(10^6)$ using wire-wound NbTi coils and commercial capacitors when the components are kept away from magnetic fields (as would be the case for a toroidal magnet surrounded by a pickup inductor) \cite{falferi1994high,nagahama2016highly}. However, they possess potentially dominant loss mechanisms in the form of normal metal joints and dielectric loss in polytetrafluoroethylene (PTFE) and formvar wire coatings. Moreover, there exists only a limited understanding of what dictates quality factor in lumped-element resonators. 

These issues motivate a systematic study to understand loss mechanisms with the aim of improving $Q$ in the 1\,kHz-100\,MHz frequency range. A program is underway to explore the use of a variety of superconductors for wires and electrodes--niobium, aluminum, titanium, tantalum, and their alloys--and crystalline capacitor dielectrics for frequency tuning--sapphire, silicon, rutile. Through careful experimental study and optimization of $Q$, superconducting radio-frequency cavities in the GHz range have demonstrated $Q$s on the order of $10^{12}$ (one million times higher than lumped-element resonators) \cite{Romanenko:2018nut}. The current goal is to improve $Q$ in the 1\,kHz-100\,MHz frequency range by a factor of $20$ to $20$ million. Resonators with improved $Q$ may be integrated into the \DMRL testbed as a demonstrator for \DMRGUT.

For resonator tuning structures, we plan to build upon the techniques developed from \DMRL and \DMRm. As described previously, the physical size of the receiver is decoupled from the search frequency; therefore, we may achieve appreciable tuning with a single resonator configuration. The resonance frequency may be tuned in the LC circuit by changing the capacitance. Simulation has indicated that we may achieve a factor of 10 change in capacitance using an insertable dielectric or a tunable electrode overlap, resulting in a factor of $\sim3$ in frequency change in a single cooldown. Such tuning mechanisms need not unacceptably degrade the $Q$; sapphire loaded superconducting cavities have demonstrated quality factors in excess of $10^{9}$ \cite{blair1985high}. For coarser tuning, inductor coil sets may be changed between cooldowns to scan different frequency ranges. In this manner, \DMRGUT can probe the desired $0.1-30\,\textrm{MHz}$ frequency range with as few as five resonator configurations.

\subsection{Active feedback}

A passive resonator is not the only option for the impedance matching network that transfers the signal to the amplifier.  In a passive lumped LC circuit, the reactance of the inductor and capacitor cancel at resonance. The power absorption from axions is thus maximized at that frequency but falls off away from resonance. To circumvent the limitation, one may instead use receivers with active elements \cite{daw2019resonant}. For example, if the capacitor is replaced with a negative inductor with negligible loss, one may cancel the reactance and achieve the maximum power absorption over a broad range of frequencies. Such wideband reactance-cancellation schemes would permit the evasion of the Bode-Fano criterion \cite{bode1945network,fano1950theoretical,Chaudhuri:2021xjd}, and they have been discussed for several decades in the engineering community such as in the context of white-light cavities in the optical regime \cite{sussman2009non,salit2010enhancement,shlivinski2018beyond}. Moreover, near-dissipationless negative inductors can be realized in the kHz and MHz regimes using Josephson Junctions \cite{clarke2006squid}. Research and development toward \DMRGUT will aim to exploit active matching elements as a potential route to realizing a GUT-scale axion search.

\section{Reach and discussion}\label{sec:reach}

\begin{figure}[h]
    \centering
    \includegraphics[width=\textwidth]{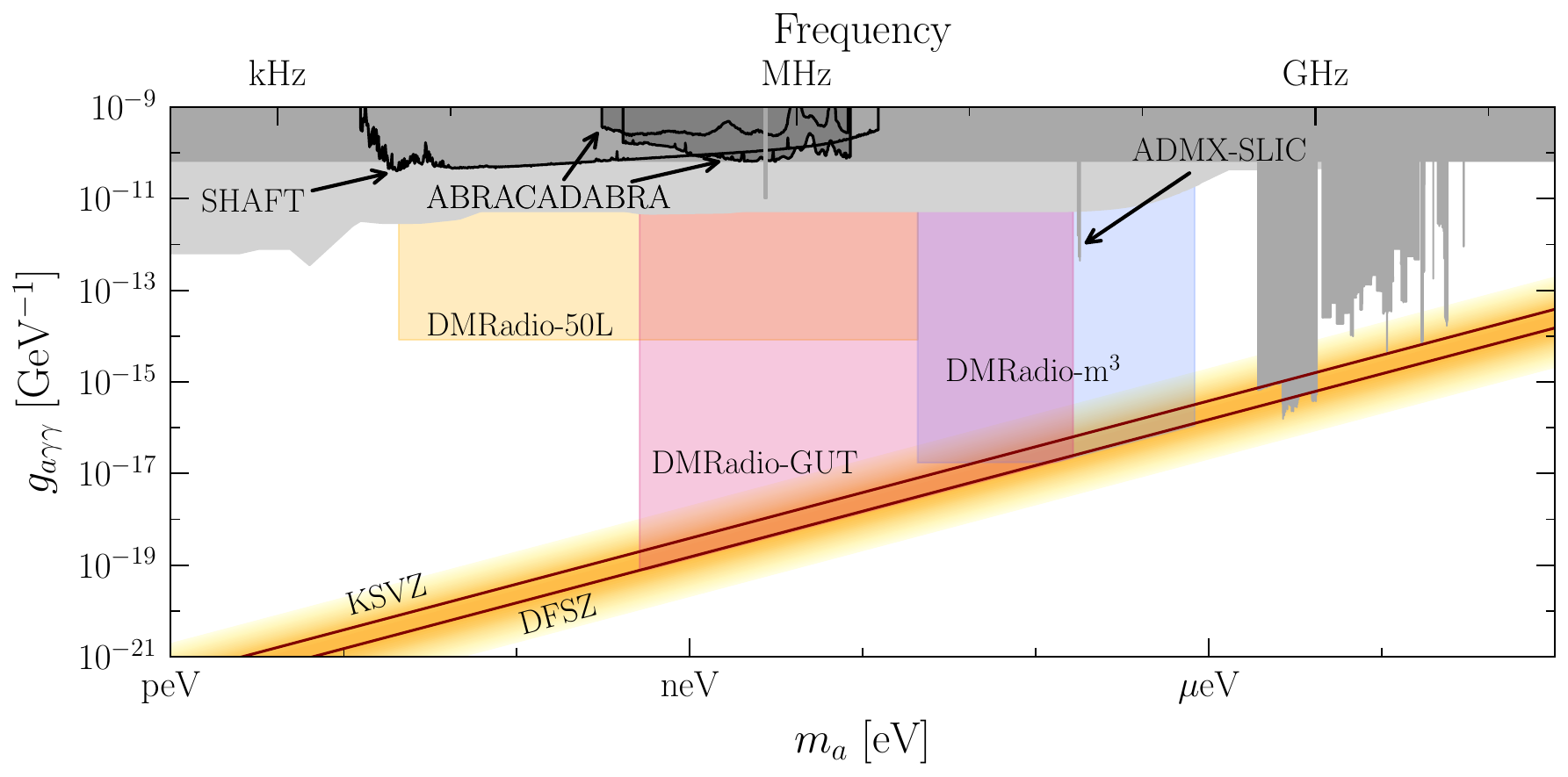}
    \caption{Projected sensitivity for \DMRGUT in pink. The total scan time to cover this reach is $\sim6$\,years, depending on R\&D outcomes. Various scenarios are outlined in \tab{paramTable}.  Existing limits are shown in grey.}
    \label{fig:reach}
\end{figure}

The program outlined above coupled with the success of current lumped-element searches suggests that the discovery of GUT-scale axions with DFSZ couplings is within reach. The predicted reach of the \DMRGUT experiment is shown in \fig{reach} using the baseline design parameters found in \tab{paramTable}.

\begin{table}[h]
    \centering
    \begin{tabular}{c c c c c c c c} \toprule
        \textbf{Scenario} & {$B_0$} & {$V$} & {$c_{PU}$} & {$Q$} & {$\eta_A$} & {$T$} & {\textbf{Scan time}} \\ \midrule
        Baseline  & 16\,T & 10\,m$^3$ & 0.1 & $20\times10^6$ & -20\,dB & 10\,mK & 6.2 years \\
        \midrule
        \specialcell{Stronger magnet +\\higher noise}  & 29\,T & 10\,m$^3$ & 0.1 & $20\times10^6$ & -5\,dB & 10\,mK & 3.2 years \\
        \midrule
        \specialcell{Lower noise +\\lower volume}  & 16\,T & 8\,m$^3$ & 0.1 & $20\times10^6$ & -25\,dB & 10\,mK & 7.3 years \\
        \midrule
        \specialcell{Higher volume +\\lower $Q$}  & 16\,T & 17\,m$^3$ & 0.1 & $2\times10^6$ & -20\,dB & 10\,mK & 10.6 years \\
        \bottomrule
    \end{tabular}
    \caption{Experimental parameters for a few different scenarios.  All scenarios reach DFSZ sensitivities at $\textrm{SNR}=3$ for $m_a\in[0.4,120]\,\textrm{neV}$ assuming $\rhoDM=0.45\,$GeV/cm$^3$.  $B_0$ is peak field strength, $V$ is volume of the pickup structure, $c_{PU}$ is the strength of the detector-axion effective current coupling (described in \sect{sensitivity}), $Q$ is the resonator quality factor, $\eta_A$ is the amplifier noise, and $T$ is the system operating temperature.  The alternative scenarios in rows 2-4 explore potential detector configurations for different R\&D outcomes.  For example, in row 2, if amplifier R\&D does not reach its goal of -20\,dB of backaction-noise reduction, increasing the magnetic field strength can counteract the sensitivity reduction and result in a comparable total scan time. The total scan times in the table are calculated from \eqn{scanRate}.
    \label{tab:paramTable}}
\end{table} 

\DMRGUT's success does not require that all R\&D tasks outlined above achieve their design goals. There can be trade-offs between the performance of each of the core systems while maintaining sensitivity to DFSZ axions with reasonable scan times. For example, if there is difficulty reducing the sensor amplifier noise far below the \SQL, then an increase in magnetic volume can compensate. A few of these scenarios are outlined in \tab{paramTable}. The most ambitious may be the stronger magnet with higher noise scenario.  However, a 29\,T peak field (20\,T \RMS) magnet is still within the R\&D goals of the greater magnet community \cite{snowmassMagnets,Awaji2017,Hahn2019}. 

A definitive search for axion DM across all masses is one of the highest priorities for particle physics and key to uncovering the particle nature of DM~\cite{kolb2018basic}. The DMRadio Program, including \DMRL, \DMRm, and \DMRGUT, is the low-frequency complement to experiments searching at higher masses. \DMRGUT is designed to address parameter space in the mass range $0.4-120\,\textrm{neV}$ ($0.1-30\,\textrm{MHz}$), probing axions at the GUT scale. It is the ultimate experiment for the lumped-element technique, harnessing key technological advancements to probe this critical parameter space.

\section{Acknowledgments}
The authors acknowledge the support of from the NSF under awards 2110720 and 2014215. \DMRm is supported by the DOE HEP Cosmic Frontier under FWP 100559. The study of resonator quality factor is supported under DOE HEP Detector R\&D (award no. DE-SC0007968). This work was supported in part by the US Department of Energy, Office of High Energy Physics program under the QuantISED program, FWP 100667. \DMRL is funded by the Gordon and Betty Moore Foundation. SLAC and UC~Berkeley gratefully acknowledge support for this work from the Gordon and Betty Moore Foundation, grant number 7941. Additional support was provided by the Heising-Simons Foundation. S.~Chaudhuri acknowledges support from the R.H. Dicke Postdoctoral Fellowship at Princeton University. C.~P.~Salemi is supported in part by the National Science Foundation Graduate Research Fellowship under Grant No.~1122374.  Y.~Kahn is supported in part by DOE grant DE-SC0015655.  B.~R.~Safdi  was  supported  in part  by  the  DOE  Early  Career  Grant  DESC0019225. P.W. Graham acknowledges support from the Simons Investigator Award no. 824870, DOE HEP QuantISED Award no. 100495, and the Gordon and Betty Moore Foundation Grant no. 7946. J.~W.~Foster was supported by a Pappalardo Fellowship.

\appendix
\section{Scan rate}\label{app:scanRateDeriv}

Here we present a sketch of a parametric derivation for the scan rate of a resonant lumped-element experiment.  This is meant to provide intuition for this kind of system rather than being a rigorous derivation for a detector setup; for the latter, see \cite{tome}.

Starting with the standard Dicke radiometer equation for the \SNR, we have
\begin{equation}
    \textrm{SNR}=\frac{P_{sig}}{P_{n}}\sqrt{\Delta\nu_{sig}\cdot\tau} \,,
\end{equation}
where $\tau$ is the integration time and $\Delta\nu_{sig}$ is the signal bandwidth.  For a low-frequency lumped-element receiver read out by a flux-to-voltage amplifier (e.g. a SQUID), such as DMRadio, we are not measuring a power at the amplifier input but a current. Moreover, the amplifier input is predominantly reactive instead of dissipative, in contrast to the amplifiers used in cavity experiments such as ADMX and HAYSTAC. The appropriate expression for SNR in DMRadio is then given by replacing power with the squared modulus of the corresponding currents,
\begin{equation}\label{eqn:Dicke}
    \textrm{SNR}=\frac{|I_{sig}|^2}{|I_{n}|^2}\sqrt{\Delta\nu_{sig}\cdot\tau} \,.
\end{equation}

From this, we would like to derive the resonator scan rate, given by
\begin{equation}
    \frac{d\nu_r}{dt} \sim \frac{\Delta\nu_{sens}}{\tau} \,.
\end{equation}
$\Delta\nu_{sens}$ is the so-called sensitivity bandwidth (also sometimes termed `visibility bandwidth'), the bandwidth over which a constant \SNR can be maintained.  $\tau$ is again an integration time, where it is now specifically the time required to achieve a given \SNR in the sensitivity bandwidth.

Rearranging \eqn{Dicke} gives us the scan rate
\begin{equation}\label{eqn:scanrate_skeleton}
    \frac{d\nu_r}{dt} \sim \frac{\Delta\nu_{sens}}{\tau}=\frac{\Delta\nu_{sens}\Delta\nu_{sig}}{\textrm{SNR}^2}\frac{|I_{sig}|^4}{|I_{n}|^4} \,.
\end{equation}
We will now step through each component of the \eqn{scanrate_skeleton} to give the scan rate in terms of experimental parameters.

The signal current is determined by the strength of the axion effective current and its coupling into the resonant circuit.  The strength of the effective current is given in \ref{eq:Jeff}, and the real magnetic field induced by it can be determined via Amp\`ere's law:
\begin{equation}
    \int\textbf{B}_{ax}\cdot d\textbf{l}=\mu_0\int\textbf{J}_{eff}\cdot d\textbf{A} \,.
\end{equation}
Using $V^{1/3}$ as a proxy for the length scale of our system and dropping numerical factors, we have
\begin{equation}
    |\textbf{B}_{ax}|\sim\mu_0|\textbf{J}_{eff}|V^{1/3} \,.
\end{equation}
Thus the magnetic energy from the axions is
\begin{equation}
    U_{ax}=|\textbf{B}_{ax}|^2V \sim \mu_0^2|\textbf{J}_{eff}|^2V^{5/3} \,.
\end{equation}

Some fraction of this energy will be coupled into the readout resonator, which we parametrize by $c_{PU}^2$, following the convention of Section VI B of \cite{tome}.  If the axion signal frequency matches the LC resonance frequency, the energy stored in the resonator is rung up according to the resonator quality factor, giving
\begin{equation}
    U_{sig} \sim c_{PU}^2Q^2U_{ax} \,.
\end{equation}
The power dissipated in the resonator is related to the signal energy and current by
\begin{equation}
    P_{sig}= 2\pi\nu U_{sig}/Q = \frac{1}{2}|I_{sig}|^{2}R \,,
\end{equation}
where $R$ is the equivalent-RLC resonator resistance. We thus arrive at
\begin{equation}
    |I_{sig}|^4\sim \frac{\nu^{2}}{R^{2}} \left(\hbar c \right)^2\left(\gagg^4\rho_{DM}^2\right)\left(c_{PU}^4Q^2B_{0}^4V^{10/3}\right) \,.
\end{equation}

The noise current has contributions from thermal noise and amplifier-added noise, the latter of which can be further subdivided into flat imprecision noise and resonator-shaped backaction noise. Within the sensitivity bandwidth, the thermal noise dominates (see \fig{biggerBW2}), so we write
\begin{equation}
    |I_n|^4\sim\frac{(k_BT)^2\Delta\nu_{sig}^2}{R^{2}} \,.
\end{equation}

The final components of our scan rate are the signal and sensitivity bandwidths.  Typically, the signal bandwidth is assumed to be identical to the axion bandwidth, but in the case where the sensitivity bandwidth is smaller than the DM bandwidth (e.g. experiments with very high $Q$), the signal bandwidth is instead equivalent to the sensitivity bandwidth,
\begin{equation}
    \Delta\nu_{sig} \sim \min\left(\Delta\nu_{sens},\frac{\nu}{Q_{axion}}\right) \,,
\end{equation}
where $Q_{axion}\sim 10^{6}$ is the characteristic quality factor governed by the DM bandwidth.

The sensitivity bandwidth is determined by optimizing the resonator-amplifier coupling.  Tuning the strength of this coupling modifies how noise is allotted between imprecision and backaction. In the limit that the physical temperature $T$ is much larger than amplifier noise temperature $T_{N}$, the optimized sensitivity bandwidth is
\begin{equation}\label{eqn:sens_opt}
    \Delta\nu_{sens}\sim\frac{n_{therm}}{\eta_{A}}\Delta\nu_r \,.
\end{equation}
Here $n_{therm}\sim k_BT/h\nu_r$ is the thermal occupation number of the LC circuit, $\eta_{A}= 2k_B T_N/h\nu_{r}$ characterizes the amplifier noise, and $\Delta\nu_r=\nu_r/Q$ is the resonator bandwidth. $\eta_{A}=1$ corresponds to the \SQL. While a complete derivation of the optimized sensitivity bandwidth is given in Appendix F of \cite{tome}, the parametrics of \eqn{sens_opt} can readily be understood from standard noise-matching concepts \cite{clerk2010introduction}. In particular, the product of the imprecision and backaction noise is proportional to $\eta_{A}^{2}$.

First, suppose that the readout is noise-matched, meaning that the imprecision and backaction noise are equal on resonance. The thermal noise is $\propto n_{therm}$ and the imprecision noise is $\propto \eta_{A}$, so on-resonance, their ratio is $\propto n_{therm}/\eta_{A}$. The sensitivity bandwidth is defined as the region where the thermal noise dominates over the imprecision noise; since the thermal noise follows a Lorentzian with respect to frequency, falling off as the inverse-squared at large detunings, the sensitivity bandwidth is then $\propto \sqrt{n_{therm}/\eta_{A}}$.

For a noise-matched amplifier, the backaction noise is $\propto\eta_{A}$. However, as we increase the coupling to the amplifier and the noise mismatch, the backaction noise increases, and the imprecision noise proportionally decreases. The backaction noise can thus be increased by a factor of $\sim n_{therm}/\eta_{A}$ up to the thermal noise $\sim n_{therm}$ without causing a degradation in on-resonance SNR.  Thus, at optimal coupling (which is a noise mismatch), the imprecision noise is $\propto \eta_{A}^{2}/n_{therm}$, and the ratio of thermal to imprecision noise is $\propto n_{therm}^{2}/\eta_{A}^{2}$. Using the same reasoning as in the noise-matched case, this results in the sensitivity bandwidth of \eqn{sens_opt}.

Rewriting gives
\begin{equation}
    \Delta\nu_{sens}\sim\frac{k_BT}{\eta_{A}Q} \,.
\end{equation}

Putting all of these pieces together, we now have a scan rate:
\begin{equation}
    \frac{d\nu_r}{dt}\sim\frac{1}{\textrm{SNR}^2}\left(\hbar c \right)^2\left(\gagg^4\rho_{DM}^2\right)\left(c_{PU}^4\nu_r^2Q^2B_{0}^4V^{10/3}\right)\left(\frac{1 }{(k_BT)^2\Delta\nu_{sig}^2}\right)\Delta\nu_{sens}\Delta\nu_{sig} \,,
\end{equation}
Assuming $\Delta\nu_{sens}>\Delta\nu_{axion}$ the rate simplifies to
\begin{equation}\label{eq:scanrate_param}
    \frac{d\nu_r}{dt}\sim
        Q_{axion}\frac{(\hbar c)^2}{\textrm{SNR}^2}\left(\gagg^4\rho_{DM}^2\nu_r\right)\left(c_{PU}^4\frac{QB_{0}^4V^{10/3}}{k_BT\eta_{A}}\right) \,,
\end{equation}
which, using precise values from the \SHM (see \cite{turner1990periodic,freese2013colloquium}) and integrated sensitivity optimization (see equations (245) and (260) and footnote 6 of \cite{tome}), becomes \eqn{scanRate} when parametrized by the baseline scenario values in \tab{paramTable}.

\end{document}